\documentclass[11pt,aps,prd,superscritptaddress,unsortedaddress,showpacs]{revtex4-1}
\def\gs{\Upsilon_\star}
\def\sun{\odot}

\usepackage[latin1]{inputenc}
\usepackage{amsmath,amsthm}
\usepackage{amsfonts}
\usepackage{amsmath}    
\usepackage{graphicx}   
\usepackage{verbatim}   
\usepackage{color}      
\usepackage{subfig}
\usepackage{subfloat}
\usepackage{hyperref}   
\usepackage{array}

\begin{document}
\title {Testing modified gravity at large distances with the HI Nearby Galaxy Survey's rotation curves}
\author {Jorge Mastache}
 \affiliation{Departamento de F\'isica Te\'orica, Instituto de F\'isica, Universidad Nacional Aut\'onoma de M\'exico, M\'exico D.F., 04510, M\'exico.}
\author{Jorge L. Cervantes-Cota}
 \email{jorge.cervantes@inin.gob.mx}
 \affiliation{Departamento de F\'isica, Instituto Nacional de Investigaciones Nucleares, Apartado Postal 18-1027, Col. Escand\'on, M\'exico DF,11801, M\'exico.}
\author {Axel de la Macorra}
 \affiliation{Departamento de F\'isica Te\'orica, Instituto de F\'isica, Universidad Nacional Aut\'onoma de M\'exico, M\'exico D.F., 04510, M\'exico.}

\begin{abstract}
  Recently a new -quantum motivated- theory of gravity has been proposed that modifies the standard Newtonian 
  potential at large distances when spherical symmetry is considered.  Accordingly, Newtonian gravity is altered by 
  adding an extra Rindler acceleration term that has to be phenomenologically determined. Here we consider a 
  standard and a power-law generalization of the Rindler modified Newtonian potential. The new terms in the gravitational 
  potential are hypothesized to play the role of dark matter in galaxies.   Our galactic model includes the mass of the 
  integrated gas, and stars for which we consider three stellar mass functions (Kroupa, diet-Salpeter, and free mass 
  model).  We test this idea by fitting rotation curves  of seventeen low surface brightness galaxies from The HI 
  Nearby Galaxy Survey (THINGS).    We find that the Rindler parameters do not perform a suitable fit  to the rotation 
  curves in comparison to standard dark matter profiles (Navarro-Frenk-White and Burkert) and, in addition, the computed parameters 
  of the Rindler gravity show a high spread, posing the model as a nonacceptable alternative to dark matter.

\end{abstract}


\maketitle

\section{Introduction}\label{sec:introduction}

The standard model of cosmology needs large amounts of dark matter and dark energy to fit data of several cosmological 
and astrophysical probes, see e.g. Refs. \cite{Frieman:2008sn,CervantesCota:2011pn}.
However, up to date there is no direct evidence that the energy density of dark matter or dark energy  is given by  particles, and dark energy 
is difficult to accommodate within the present understanding of quantum field theory. Therefore, alternatives to both dark matter and dark energy 
are abundant in the literature, see Ref. \cite{Clifton:2011jh} for current approaches.

Recently a new -quantum motivated-  theory of gravity has been proposed that modifies the standard Newtonian potential at large distances
when spherical symmetry is considered \cite{Grumiller:2010bz}.  The formalism incorporates a Rindler acceleration term
that could serve to explain the galactic dynamics without the presence of a dark matter halo \cite{Grumiller:2011gg}. This seems
plausible since  a rough estimate of the Rindler acceleration is of the order $\sim 10^{-9} {\rm cm}/{\rm s}^2$, a value similar to
MOND's acceleration that has been proven to successfully explain rotation curves without a dark matter halo \cite{MOND1,MOND2,Famaey:2011kh}; other
applications to this model have been computed elsewhere \cite{Carloni:2011ha,Iorio:2011zu,Li:2011ur,Sultana:2012zz,Culetu:2012yh}.   The
aim of the present work is to test this idea when applied to rotation curves of spirals.    To perform this task, we use the HI Nearby Galaxy Survey (THINGS), which collects
high spectral resolution data revealing extended measurements of  gas rotation velocities and circular baryonic matter
trajectories \cite{Walter:2008wy}. Given these properties it is adequate to test the new gravity model
with THINGS, which has been used to test  different core/cusp mass profiles. For disk-dominated galaxies
the core and cusp profiles fit equally well, however for low surface brightness galaxies there is a clear preference for core
profiles over the cuspy models \cite{deBlok:2008wp}. Analyses of different high resolution datasets have confirmed this tendency in past recent years
\cite{SpGiHa05,KuMcBl08,Salucci:2010pz,Gentile:2004tb,Salucci:2007tm}.  In this work, we present a  study of the rotation curves using the
new gravity model, taking into account the contribution from gas
and the three stellar disk models (Kroupa, diet-Salpeter, free mass).  By  fitting the models to the data we find that  although the fits are
achievable for the considered galaxies, in
many cases they show high  $\chi^{2}_{\rm red}$ values, and a high spread in the Rindler parameters ($a$, $n$).
Furthermore, the standard dark matter profiles [Navarro-Frenk-White (NFW) and Burkert] do a better job in the fittings to rotation curves.    A very recent 
work  \cite{Lin:2012zh} considers  the same problem and it uses the same eight (of our seventeen) galaxies for the standard Rindler 
model ($n=1$), concluding that for six galaxies their results tend to converge to a single Rindler acceleration parameter. In our 
case, we observe this evidence too but when one takes into account more galaxies or other stellar galactic models their conclusions 
do not hold.

This work is organized as follows:  In Sec. \ref{Grumiller_theory} we review the basic concepts of galactic dynamics when applied to the
Rindler acceleration as implied from Ref.  \cite{Grumiller:2010bz}. In Sec. \ref{sample} we describe the THINGS galactic sample and
in Sec. \ref{MaMo} explain how the gas is treated and the stellar mass models considered, implying different galactic models.  Then in
Sec. \ref{results} we analyze the results yielded from the minimization procedure as best fits of the different models. Finally, the
conclusions are drawn in Sec. \ref{conclusions}.

\section{Modified gravity at large distances} \label{Grumiller_theory}

In Grumiller's approach \cite{Grumiller:2011gg} the effective potential of a point mass ($M_i$), without angular momentum contribution,  is:
\begin{equation}\label{potunit}
    \phi_i(x_i) = - G \frac{M_i}{x_i} + a \,  x_i,
\end{equation}
where  $x_i = |\vec{x_i}| = |\vec{r} - \vec{r_i}^\prime|$, $\vec{r}$ is an arbitrary point, $r^\prime$ are the particle 
coordinates, and \emph{a} is a universal constant, the Rindler acceleration.

For a smooth matter distribution with  spherical symmetry,  one has
\begin{equation}\label{intpot}
    \phi (r) = - G \int \! \frac{\rho(r\prime)}{|\vec{r}-\vec{r\prime}|} \, \mathrm{d^3} r\prime + a \, |\vec{r}| ,
\end{equation}
where $\rho(r)$ is the density profile at radius $r$.   The corresponding rotation velocity yields
\begin{equation}\label{vel1}
    v^2 (r) = v_{\rm N}^2(r) + v_{R}^{2}(r),
\end{equation}
where the subindex ``N" stands for the Newtonian contribution and the Rindler circular velocity is given by
\begin{equation}\label{aRind}
v_{R}^{2}(r) \equiv a \, |\vec{r}| .
\end{equation}
A similar approach to  Eq. (\ref{vel1}) was presented in Ref. \cite{Mannheim:1996rv} to fit rotation curves
within a conformal gravity theory; see Refs. \cite{Mannheim:2005bfa,Mannheim:2010ti} for further developments of this model.  As emphasized in
Ref. \cite{Grumiller:2011gg}, the Rindler term constitutes a rough model  which casts doubts on the description of rotation curves with such a linear
growing of the velocity with the radius. It was therefore suggested to consider some kind of $r$-dependent term in the acceleration to test a more general
Rindler hypothesis. Accordingly, we will assume below a power-law dependence that is a next step in complexity that adds an extra parameter, the exponent $n$  
in Eq. (\ref{aRind.gen}) below.   One may wonder how many extra parameters one might need to fit the data, and whether this represents a better fit than 
standard dark matter models. For instance,  typical dark matter halos, e.g.  NFW's  
\cite{Navarro:1995iw,Navarro:1996gj} or  Burkert's \cite{Burkert:1995yz} profiles, have 
two free parameters, a scale radius and density parameter.  Then, from a phenomenological point of view, having two parameters in a dark matter  
alternative seems not too demanding, since it has the same number of free parameters to perform the fits. At the end, the best fits of the alternative model 
should be compared with the ones of the standard dark matter models, as it will be done  at the end of the results section below.    Let us remark that  
in standard dark matter profiles one expects some variation of the two free parameters from galaxy to galaxy, but  this does not happen in Grumiller's models 
since the two free parameters are constants in every galaxy; $a$ is fundamental constant of nature and exponent $n$ is also a constant, yet to be 
determined. Therefore, the challenge to fit is bigger in Grumiller's models, and this is a price paid to modify gravity laws. 

The main luminous components in a typical spiral galaxy are gas, stars, and a bulge, which are orbiting around the galactic center, and normally a 
dark matter halo is assumed to account for the high speed of the rotation curves. In the model used in the present work there exists no dark matter, but 
instead we use a Rindler term to explain the velocity distribution of stars and gas. Accordingly, the model developed here considers spherical symmetry 
and physical quantities will be independent of the polar angle. Therefore, the Newtonian contribution from stars ($\star$) and gas ($G$) can be thought 
as given by a spherized disk. The contribution from stars is then given by the Freeman disk
\cite{Fr70,BaSo80}:
\begin{equation} \label{rho-stars}
  \rho_{\star} (r) = \frac{M_d}{2 \pi r_{d}^{2}}  \, e^{-r/r_{d}} ,
\end{equation}
where $M_d$ is the mass of the disk and  $r_d$ its radius. Thus, the rotation curve contribution from stars with a standard Newtonian dynamics yields \cite{BiTr08}
\begin{equation}\label{v2_stars}
  v_{\star}^2(r)  = \frac{G M_d }{2 r_{d}} \left(\frac{r}{ r_{d}}\right)^{2}
  \left[ {\rm I_{0}}\left(\frac{r}{2 r_{d}}\right) {\rm K_{0}} \left(\frac{r}{2 r_{d}}\right) -
  {\rm I_{1}}\left(\frac{r}{2 r_{d}}\right) {\rm K_{1}} \left(\frac{r}{2 r_{d}}\right) \right] \, ,
\end{equation}
where the functions I and K are the modified Bessel functions; more details on the treatment of stars are given in Sec. \ref {MaMoEsDi}.  On the other 
hand, the gas contribution ($v^2_{G}$) is computed by integrating its surface brightness as in standard Newtonian lore, as we will explain in Sec. \ref{MaMoGa}.

Bringing together all contributions to the total ($T$) rotation curve
and including a generalized Rindler (GR) term,
\begin{equation}\label{vel2}
    v_{T}^2 (r) = \gs v^2_{\star} + v^2_{G} + v_{GR}^{2}(r) \, ,
\end{equation}
where $\gs$ is the  mass-to-light ratio and
\begin{equation}\label{aRind.gen}
 v_{GR}^{2}(r) \equiv a |\vec{r}|^n .
\end{equation}
The case $n=1$ is the original model of modified gravity at large distances \cite{Grumiller:2011gg}, as in Eqs. (\ref{intpot}, \ref{vel1}).
The new free parameters of the model of galactic rotation curves are $a$ and $n$, and they have to be determined by observations, instead of the
two free parameters of the standard cold dark matter galactic profile, such as in NFW \cite{Navarro:1995iw,Navarro:1996gj}, or alternative 
the Burkert  \cite{Burkert:1995yz}, pseudoisothermal, or bound dark matter (BDM) \cite{delaMacorra:2009yb} profiles; for a comparison of these profiles
see  Refs. \cite{delaMacorra:2011df,Mastache:2011cn}.    To extract information for the Rindler parameters, as an
input we will need the total, observational rotation curve $v_{T}$, the rotation curve of the stellar component $v_\star$, and the gas
component $v_G$.  Following, we describe the observational data used and the models behind each component.

\section{The THINGS galactic sample}\label{sample}

We make use of the HI data provided by THINGS, which possess high resolution velocity fields of rotation curves that are ideal to test new dark matter profiles  \cite{Mastache:2011cn,delaMacorra:2011df}  or, as it is our present case, to test new gravitational theories.  THINGS galaxies have an observing data sample of 34 nearby galaxies containing a large range of luminosities and Hubble types, but we limit our sample to 17
low luminous (early type and dwarf) galaxies with smooth, symmetric and extended to large radii rotation curves and small or
no bulge, see Table \ref{tab:things1}. These properties can provide a good estimate to alternative models because their underlying dynamics is believed to be dominated by dark matter --or new alternative approaches-- over all other components at all radii.
For technical details and systematic effects treatment of the observations of the THINGS sample refer to Refs.
\cite{Walter:2008wy,Trachternach:2008wv,Oh:2008ww}. For a complete analysis of its rotation curves, see Refs. \cite{deBlok:2008wp,Mastache:2011cn}.

In addition to the rotation curves extracted from THINGS data and we use the $3.6 \ {\rm \mu m}$ data from Spitzer Infrared Nearby Galaxies Survey (SINGS) \cite{Kennicutt:2003dc}. We follow the analysis of  Refs. \cite{deBlok:2008wp,McGaugh:2006vv} for the sample considered here.

\begin{table}[h]
\centering
\scriptsize{
\begin{tabular}{lcccc}
\multicolumn{5}{c}{THINGS DATA} \\
 \hline  \hline
               &            &            & DIET-SALPETER &     KROUPA \\
       \multicolumn{1}{c}{Galaxy}  & Distance & $R_d $ & $\log_{10} M_{star}$ & $\log_{10} M_{star}$ \\
       \multicolumn{1}{c}{(1)} & (2) & (3) & (4) & (5) \\
       \hline
        N 925 &        9.2 &      3.30 &      10.01 &       9.86 \\
       N 2366 &        3.4 &      1.76 &       8.41 &       8.26 \\
       N 2403 &        3.2 &       1.81 &       9.67 &       9.52 \\
       N 2841 &       14.1 &    4.22 &      11.04 &      10.88 \\
       N 2903 &        8.9 &    2.40 &      10.15 &         10.0 \\
       N 2976 &        3.6 &   0.91 &       9.25 &        9.10 \\
       N 3031 &        3.6 &     1.93 &      10.84 &      10.69 \\
       N 3198 &       13.8 &       3.06 &      10.45 &       10.30 \\
       N 3521 &       10.7 &    3.09 &      11.09 &      10.94 \\
       N 3621 &        6.6 &       2.61 &      10.29 &      10.14 \\
       N 4736 &        4.7 &       1.99 &      10.27 &      10.12 \\
       N 5055 &       10.1 &     3.68 &      11.09 &      10.94 \\
       N 6946 &        5.9 &       2.97 &      10.77 &      10.62 \\
       N 7331 &       14.7 &    2.41 &      11.22 &      11.07 \\
       N 7793 &        3.9 &     1.25 &       9.44 &       9.29 \\
        I 2574 &        4.0 &    2.56 &       9.02 &       8.87 \\
       D 154  &        4.3 &       0.72 &       7.42 &       7.27 \\
    \end{tabular}
    }
  \caption{\footnotesize{Sample of THINGS late-type and dwarf galaxies as presented in Walter {\it et al}
  \cite{Walter:2008wy}. Columns: (1) Galaxy's name. (2) Distance to the galaxy in Mpc. (3) Characteristic radius of
  the stellar disk in kpc as given in Ref. \cite{deBlok:2008wp}. (4) Logarithm of the stellar mass disk when considering
  the diet-Salpeter IMF in solar masses ($M_\sun$), and (5) Logarithm of the stellar mass disk when considering the
  Kroupa IMF ($M_\sun$).
  }}
  \label{tab:things1}
\end{table}

\section{Treatment of gas and stars} \label{MaMo}
Our galactic model includes the two main observable components of a spiral galaxy: thin gaseous disk and  a
thick stellar disk. In most cases the stellar disk can be well described by a single exponential disk,  given
by Eq. (\ref{rho-stars}). When necessary, in a small number of galaxies we have considered an additional central
component, a bulge, containing a small fraction of the total luminosity of the galaxy, as described by Ref. \cite{deBlok:2008wp}.

Following, we describe in more detail the treatment for the stellar component in Sec. \ref{MaMoEsDi}, as well for the gas component in Sec. \ref{MaMoGa}.

 \subsection{Stellar distribution  \label{MaMoEsDi}}
To model the stellar disk when the $\gs$ is assumed to be constant at all radii, we use the approximation of a radial exponential 
profile of zero thickness, the Freeman disk \cite{Fr70, BaSo80}, since the disk vertical scale height does not change appreciably 
with radius and the correction to the velocity is around 5\% in most cases \cite{Burlak:1997}. Then, the central surface density is 
given by  Eq. (\ref{rho-stars}), where $r_d$ is the scale 
length of the disk and $\Sigma_0 \equiv \frac{M_d}{2 \pi r_{d}^{2}} $ is the central surface density with units [$M_\sun {\rm pc}^{-2}$]. These
two parameters are obtained first by fitting the observed surface brightness profile, extracted from the SINGS images at the $3.6 \mu$m 
band and synthesized by Ref. \cite{deBlok:2008wp}, to the linear formula $\mu(R) = \mu_0 + 1.0857 \, r/r_{d}$, where $\mu_0$
is the central surface brightness given in observational units (mag arcsec$^{-2}$), and $\mu_0$ and the surface brightness are related
by a simple change of units.  We get the surface density thanks to the mass-to-light ratio $\gs$, the standard additional free parameter
in the mass model, introduced because we generally can only measure the distribution of the light instead of the mass.
Stellar disks sometimes show radial color gradients, and it is believed that this provides an indication of stellar population between the inner and
outer regions of a galaxy and produce $\gs$ gradients between these two regions of the disk \cite{Taylor:2005sf}. We take the $\gs$ as a function
of the radius in order to consider the different star's contribution as it depends on the region that we are analyzing.

When we estimate the Rindler parameters $(a,n)$, $\gs$ is an important source of uncertainty, because these parameters 
are degenerate through Eq. (\ref{vel2}). However, because stars have a major contribution near the center of the galaxy and 
the Rindler acceleration should contribute at large distance, we expect that $\gs$ does not significantly contribute  to the 
uncertainties of the Rindler parameters.  In the results section, we will show this is the case for most of the galaxies.

The $\gs$ has been modeled, e.g. in Salpeter \cite{Salpeter:1955it}, Kroupa \cite{Kroupa:2000iv}, and Bottema \cite{Bottema:1997qe}, but the
precise value for an individual galaxy is not well known and depends on extinction, star formation history, initial mass function (IMF), among
others. Some assumptions have to be made respect to $\gs$ in order to reduce the number of free parameters in the model.
In our case we studied three different $\gs$ models:  (i) we obtained  $\gs$ from the galaxy colors as predicted
by spectrophotometric models with Kroupa \cite{Kroupa:2000iv} initial mass function (IMF) which is based on stellar population studies in the Milky Way
and it yields low disk masses that minimizes the baryonic contribution;  (ii) we determine $\gs$ through a diet-Salpeter \cite{Salpeter:1955it}, in which stellar 
mass population syntheses have proved \cite{Bell:2000jt} to maximize the disk mass contribution; and  (iii) we assume the standard method in which the
stellar $\gs$ is a model-independent free parameter. The resulting mass models reproduce well the rotation curves, and the Rindler parameters are
derived within a reasonable uncertainty.

In our fittings of Kroupa and diet-Salpeter models we considered radial color gradients, but for the free mass model we consider
a $\gs$ as an unknown, constant parameter to be determined.  In all three models we took into account the bulge, when present, to
contribute  to  the stellar disk.

The contribution of the atomic gas is considered in all three stellar Kroupa, diet-Salpeter, and free mass models, and it is
briefly described as follows.

 \subsection{Neutral gas distribution \label{MaMoGa}}

 For the gas we assumed an infinitely thin disk in order to compute the corresponding rotation curve. For more technical details we
 refer to  Ref. \cite{deBlok:2008wp}. We point out that the case of a disk with sufficient central depression in the mass distribution
 can yield a net force pointing outwards, and this generates an imaginary rotation velocity and therefore a negative $v_G^2$. An
 imaginary velocity is just a reflect of the effective force of a test particle caused by a nonspherical mass distribution with a
 depression mass in the center. We have not included the contribution of the molecular gas since its surface density is only
 a few percent of that of the stars, therefore its contribution is reflected in a small increase in $\gs$ \cite{Portas:2009}.

\section{Results}\label{results}

In order to constrain the Rindler modification of gravity we considered two scenarios.  First, we treated the original Rindler model with $n=1$ and,   second,  the power-law dependence Rindler model, as suggested in Ref. \cite{Grumiller:2010bz}, with two free parameters $(a, n)$, as explained in
Sec. \ref{Grumiller_theory}.  For both cases we considered the Kroupa minimal disk and the diet-Salpeter maximum disk.  In the third place, we present the
fittings for Rindler models with $n=1$ and $n\neq1$ using the Free $\gs$ stellar mass function.

We make use of  the observed rotation curve, stellar, and gas components as an input for the numerical code, in order to obtain the Rindler
parameters. To fit the observational velocity curve with the theoretical model we employ the $\chi^2$ goodness-of-fit test ($\chi^2$ test), that
tells us how ``close" are the theoretical to the observed values. In general the $\chi^2$ test statistics are of the form 
\begin{equation}
   \chi^2 = \sum_{i=1}^n \left(\frac{v_{{\rm obs}_i}-v_{{\rm model}_i}(r, a, n)}{\sigma_i}\right)^2,
\end{equation}
where $\sigma$ is the standard deviation, and $n$ is the number of observations.  One defines the
reduced $\chi^{2}_{\rm red} \equiv \chi^2/({\rm n}-p-1)$, in which n is the number of observations and $p$ is the number of fitted  parameters.

\subsection{Fitting the standard ($n=1$) Rindler model with Kroupa and diet-Salpeter IMF} \label{fitting_n=1}
We considered the original Rindler model with $n=1$ and proceed to fit the parameter $a$ for the Kroupa minimal disk and
diet-Salpeter maximum disk with a $\chi^2$ test. When we minimize $\chi^2$ we use different methods (differential evolution, Nelder-Mead, and simulated
annealing) to guarantee not to be in a local minimum. At the same time, we assume priors on the free parameter in order to have
values greater than or equal to zero to obtain physical reasonable values.  The results are presented in Table \ref{tab:things2}, where
the $\chi^{2}_{\rm red}$ values are given. Notice that the different stellar mass models do not significantly change the
determined value of the Rindler acceleration for most of the galaxies.
The uncertainties in the rotation velocity are reflected in the uncertainties in the model
parameters but in general these are small. One observes some spread in the values for $a$ (in units of $\rm{cm}/{\rm s}^2$), ranging
from   $ 0.93^{+ 0.01}_{- 0.44}$ for N 2366 to $9.57^{+ 0.06}_{- 0.06}$ for N 2841 in Kroupa's model, to account for a difference of
an order of magnitude, but the uncertainties are small and
they do not account for such a difference.  In addition to this discrepancy,  the
fits to some of the galaxies present very high  $\chi^2_{\rm red}$ values, that speaks for a poor fitting. In the diet-Salpeter mass model, $a$
varies in a similar fashion, given $ 0.62^{+ 0.10}_{- 0.10}$ for N 3031 and $  7.79^{+ 0.06}_{- 0.06}$ for  N 284, and again the
goodness-of-fit test
is not satisfactory for most  of the galaxies.  By comparing both fits, Kroupa did better in 9 (out of 17) cases and diet-Salpeter in 8.
None of the fits with $n=1$ had a $\chi^2_{\rm red} \le 1$.

\setlength{\extrarowheight}{3pt}
\begin{table}[h]
  \centering
    \begin{tabular}{r|rrr|rrr}
                                                          \multicolumn{ 7}{c}{Rindler $n=1$} \\
                                                          \hline
               &          \multicolumn{ 3}{c|}{Kroupa} &   \multicolumn{ 3}{c}{diet-Salpeter} \\
        Galaxy & $a$ & $a$ $[\frac{\rm{cm}}{{\rm s}^2}]$ &     $\chi^{2}_{\rm red}$ &$ a$ & $a$ $[\frac{\rm{cm}}{{\rm s}^2}]$ &     $\chi^{2}_{\rm red}$ \\
        \hline
        D 154 & $  358.45^{+   5.45}_{-   5.49}$ & $ 1.16^{+ 0.02}_{- 0.02}$ &       2.10 & $  355.64^{+   5.45}_{-   5.48}$ & $  1.15^{+ 0.02}_{- 0.02}$ &       2.03 \\
        I 2574 & $  297.35^{+   6.91}_{-   6.97}$ & $ 0.96^{+ 0.02}_{- 0.02}$ &       4.74 & $  272.61^{+   6.91}_{-   6.97}$ & $  0.88^{+ 0.02}_{- 0.02}$ &       5.53 \\
       N 2366 & $  285.45^{+   1.94}_{- 134.51}$ & $ 0.93^{+ 0.01}_{- 0.44}$ &       3.48 & $  285.45^{+   2.66}_{-  99.97}$ & $  0.93^{+ 0.01}_{- 0.32}$ &       3.29 \\
       N 2403 & $ 1258.65^{+   5.77}_{-   5.78}$ & $ 4.08^{+ 0.02}_{- 0.02}$ &      11.80 & $ 1225.23^{+   5.75}_{-   5.76}$ & $  3.97^{+ 0.02}_{- 0.02}$ &       9.29 \\
       N 2841 & $ 2952.36^{+  19.45}_{-  19.50}$ & $ 9.57^{+ 0.06}_{- 0.06}$ &      76.10 & $ 2405.18^{+  18.91}_{-  18.96}$ & $  7.79^{+ 0.06}_{- 0.06}$ &      44.30 \\
       N 2903 & $ 1985.81^{+  13.91}_{-  13.96}$ & $ 6.44^{+ 0.05}_{- 0.05}$ &      71.80 & $ 1998.74^{+  13.92}_{-  13.97}$ & $  6.48^{+ 0.05}_{- 0.05}$ &      74.10 \\
       N 2976 & $ 1000.00^{+   2.64}_{- 723.12}$ & $ 3.24^{+ 0.01}_{- 2.34}$ &       8.05 & $  698.04^{+  44.46}_{-  44.80}$ & $  2.26^{+ 0.14}_{- 0.15}$ &      10.70 \\
       N 3031 & $ 2617.86^{+  32.68}_{-  32.77}$ & $ 8.48^{+ 0.11}_{- 0.11}$ &       9.54 & $  191.44^{+  29.52}_{-  29.62}$ & $  0.62^{+ 0.10}_{- 0.10}$ &      22.80 \\
       N 3198 & $  632.49^{+   5.85}_{-   5.87}$ & $ 2.05^{+ 0.02}_{- 0.02}$ &      14.00 & $  567.72^{+   5.78}_{-   5.80}$ & $  1.84^{+ 0.02}_{- 0.02}$ &      11.50 \\
       N 3521 & $  684.59^{+  36.21}_{-  36.59}$ & $ 2.22^{+ 0.12}_{- 0.12}$ &       6.14 & $  562.41^{+  35.10}_{-  35.50}$ & $  1.82^{+ 0.11}_{- 0.12}$ &       7.54 \\
       N 3621 & $  889.79^{+   6.98}_{-   7.00}$ & $ 2.88^{+ 0.02}_{- 0.02}$ &      11.20 & $  777.46^{+   6.86}_{-   6.88}$ & $  2.52^{+ 0.02}_{- 0.02}$ &       3.29 \\
       N 4736 & $ 1005.43^{+  27.96}_{-  28.11}$ & $ 3.26^{+ 0.09}_{- 0.09}$ &      20.50 & $  420.13^{+  27.07}_{-  27.24}$ & $  1.36^{+ 0.09}_{- 0.09}$ &       8.74 \\
       N 5055 & $  559.90^{+   8.71}_{-   8.74}$ & $ 1.81^{+ 0.03}_{- 0.03}$ &       5.09 & $  399.76^{+   8.30}_{-   8.34}$ & $  1.30^{+ 0.03}_{- 0.03}$ &      15.80 \\
       N 6946 & $ 1650.37^{+  16.51}_{-  16.55}$ & $ 5.35^{+ 0.05}_{- 0.05}$ &       1.29 & $  908.03^{+  16.12}_{-  16.17}$ & $  2.94^{+ 0.05}_{- 0.05}$ &       4.42 \\
       N 7331 & $ 1938.02^{+  24.45}_{-  24.55}$ & $ 6.28^{+ 0.08}_{- 0.08}$ &       8.65 & $ 1505.92^{+  23.90}_{-  24.00}$ & $  4.88^{+ 0.08}_{- 0.08}$ &      29.60 \\
       N 7793 & $ 1720.65^{+   1.57}_{- 358.31}$ & $ 5.58^{+ 0.01}_{- 1.16}$ &      10.50 & $ 1688.18^{+  37.17}_{-  15.34}$ & $  5.47^{+ 0.12}_{- 0.05}$ &       5.52 \\
        N 925 & $  612.29^{+  14.40}_{-  14.39}$ & $ 1.98^{+ 0.05}_{- 0.05}$ &       3.74 & $  453.50^{+  14.26}_{-  14.36}$ & $  1.47^{+ 0.05}_{- 0.05}$ &       5.72 \\
    \end{tabular}
  \caption{\footnotesize{Fits for the fixed power $n = 1$  using the Kroupa and diet-Salpeter mass models. In columns (2) and (5) the Rindler acceleration
  has units of ${\rm \frac{km^2}{{\rm s}^2 kpc}}$ and in columns (3) and (6) in  units ${\rm \frac{\rm{cm}}{s^2}}\times10^{-9}$.}}
  \label{tab:things2}
\end{table}

\subsection{Fitting the power-law  generalized Rindler model with Kroupa and diet-Salpeter IMF} \label{fitting_n}

We now considered a Rindler model with power-law dependence, as suggested in Ref. \cite{Grumiller:2010bz},  with two free parameters $(a, n)$ and 
we used the same stellar and gas models. The results are shown in Table \ref{tab:things3}.  Now, the values of $a$ are given in the 
units ($\frac{m^2}{{\rm s}^2 kpc^n}$) in consistency with the given $n$ values; one could extract an acceleration parameter here if one 
defines $a \, r^{n} \equiv a_{\rm new} r \,  (r/r_{\rm new})^{n-1}$, but we would only add an extra parameter ($r_{\rm new}$) that  is completely 
degenerated with $a_{\rm new}$.

\begin{figure}[h]
  \includegraphics[width=13cm]{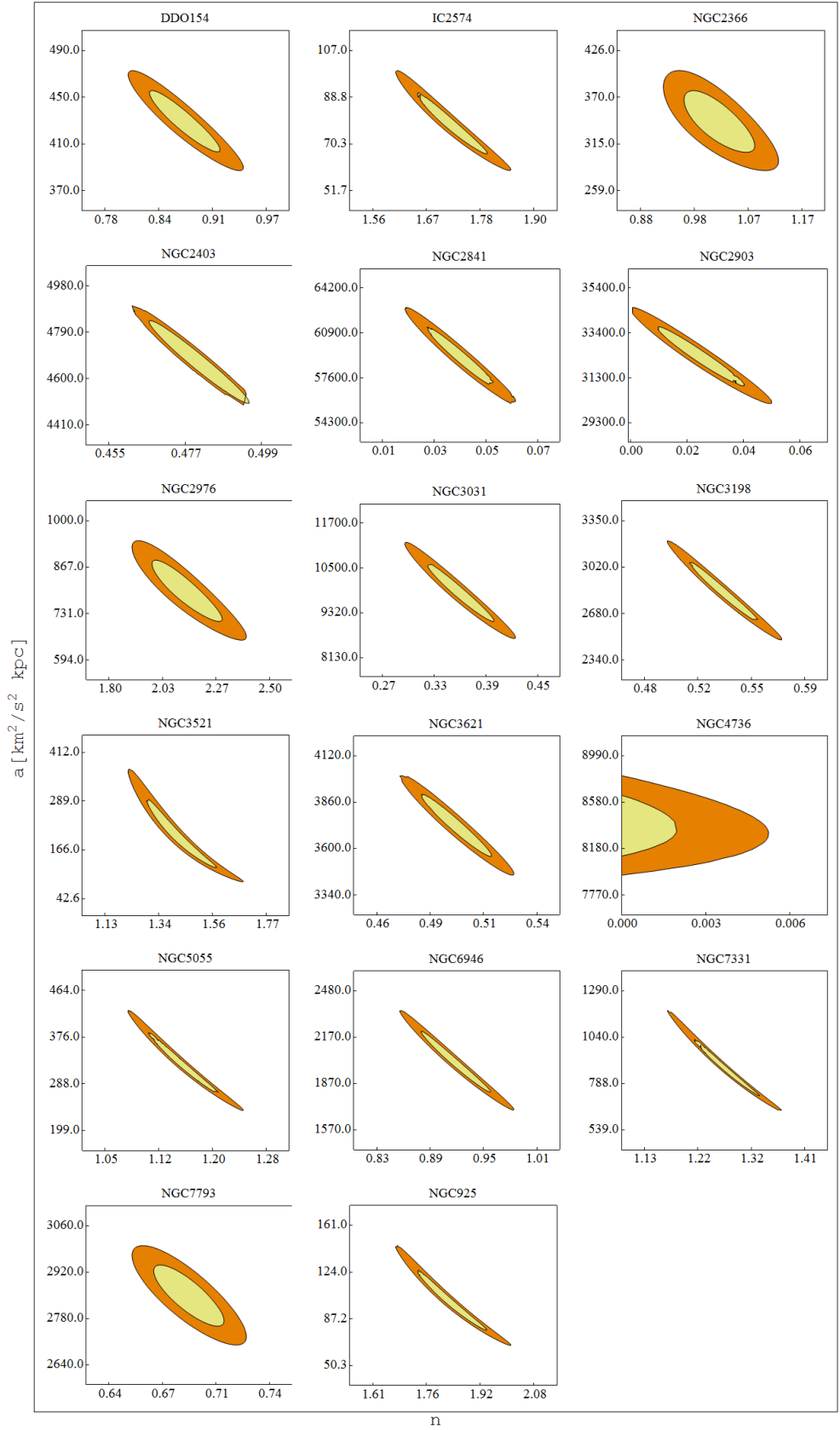}\\
  \caption{\footnotesize{2D likelihood contours between the $a$ and $n$, corresponding to $1\sigma$ and $2\sigma$, for the Kroupa IMF model.}}
  \label{tab:Kroupa}
\end{figure}

\begin{figure}[h]
  \includegraphics[width=13cm]{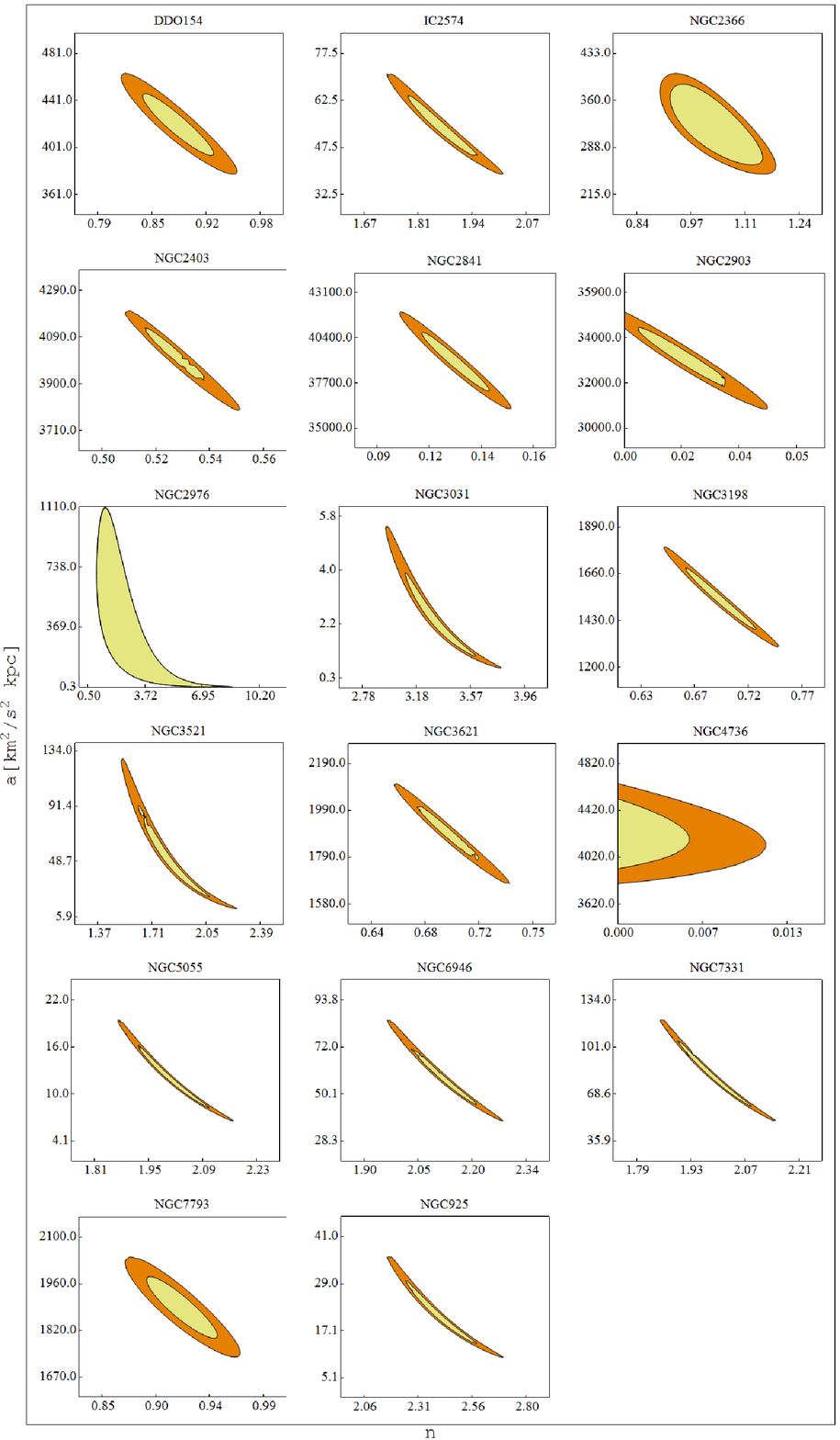}\\
  \caption{\footnotesize{2D likelihood contours between the $a$ and $n$, corresponding to $1\sigma$ and $2\sigma$, for the diet-Salpeter IMF model.}}
  \label{tab:Salpeter}
\end{figure}

The results of the fittings show a broader spread in the $a$ values.  Again, the different stellar mass models do not significantly change the determined 
value of the Rindler acceleration or power-law exponent for most of the galaxies.
By considering the Kroupa model one has $a = 77.96^{+   1.74}_{-   1.75}$ for  IC2574, and  $a = 32294.30^{+ 211.39}_{- 212.01}$ for N 2903, accounting for 2 orders of magnitude in difference.  For this IMF model, the variation on $n$ ranges from values less than $0.002$ for N 4736 to $ 2.14^{+ 0.045}_{- 0.044}$ for N 2976, that yields a 3 orders of magnitude difference.  Furthermore, the overall fitting results
are again not quite satisfactory since  some $\chi^{2}_{\rm red}$ values are high.   The analysis of the diet-Salpeter mass model shows a similar behavior:
 the ``acceleration" parameter ranges from $ 2.12^{+   0.12}_{-   0.12}$  for N 3031  to $ 4216.82^{+ 176.24}_{- 176.98}$  for N 4736, whereas the power-law
 exponent goes from very small values ($< 0.003$) for N 4736  to $ 3.43^{+ 2.196}_{- 0.001}$ for  N 2976.  For these IMF models the deviations of the
 Rindler parameters are a few orders of magnitude different.  In Figs. \ref{tab:Kroupa} and \ref{tab:Salpeter}  we present the contour plots of the Rindler
 parameters $(a,n)$ for 1$\sigma$ and 2$\sigma$ for both stellar models.

By comparing both Kroupa  and diet-Salpeter fits, the former did much better in 14 (out of 17) cases and the later in only 3.  However, the
goodness-of-fit test does not render acceptable results since some galaxies present  very high  $\chi^2_{\rm red}$ values, and only three of
them have fits with $\chi^2_{\rm red} < 1$ (all for the Kroupa IMF model).   Moreover, by considering only these three best fits, the most
favored values of $a$ and $n$ are excluded from each other within various standard deviations.

 Now, by comparing the best fits of Tables
\ref{tab:things2} and  \ref{tab:things3}, it is clear that the generalized Rindler model fits better for all the galaxies.

\setlength{\extrarowheight}{3pt}
\begin{table}[h]
  \centering
    \begin{tabular}{r|rrr|rrr}
                                                         \multicolumn{ 7}{c}{Rindler $n\neq1$} \\
                                                         \hline
               &          \multicolumn{ 3}{c|}{Kroupa} &   \multicolumn{ 3}{c}{diet-Salpeter} \\
        Galaxy & $a$ &       $   n$ &       $\chi^{2}_{\rm red}$ & $a$ &          $n$ &       $\chi^{2}_{\rm red}$ \\
        \hline
        D 154 & $  429.25^{+   6.52}_{-   6.56}$ & $ 0.87^{+ 0.010}_{- 0.010}$ &       1.77 & $  420.27^{+   6.43}_{-   6.47}$ & $ 0.88^{+ 0.010}_{- 0.010}$ &       1.75 \\
        I 2574 & $   77.96^{+   1.74}_{-   1.75}$ & $ 1.72^{+ 0.011}_{- 0.011}$ &       0.98 & $   53.57^{+   1.28}_{-   1.29}$ & $ 1.87^{+ 0.011}_{- 0.011}$ &       1.32 \\
       N 2366 & $  340.76^{+  16.15}_{-  16.42}$ & $ 1.02^{+ 0.029}_{- 0.028}$ &       3.27 & $  285.45^{+  18.13}_{-  11.09}$ & $ 1.13^{+ 0.100}_{- 0.009}$ &       3.23 \\
       N 2403 & $ 4667.30^{+  20.97}_{-  21.01}$ & $ 0.48^{+ 0.002}_{- 0.002}$ &       2.27 & $ 4000.25^{+  18.46}_{-  18.50}$ & $ 0.53^{+ 0.002}_{- 0.002}$ &       2.55 \\
       N 2841 & $ 59071.70^{+ 333.82}_{- 334.49}$ & $ 0.04^{+ 0.002}_{- 0.002}$ &       0.90 & $ 38957.50^{+ 262.73}_{- 263.27}$ & $ 0.13^{+ 0.002}_{- 0.002}$ &       1.23 \\
       N 2903 & $ 32294.30^{+ 211.39}_{- 212.01}$ & $ 0.03^{+ 0.002}_{- 0.002}$ &       7.99 & $ 33026.10^{+ 215.04}_{- 215.67}$ & $ 0.02^{+ 0.002}_{- 0.002}$ &       8.34 \\
       N 2976 & $  796.89^{+  25.89}_{-  26.15}$ & $ 2.14^{+ 0.045}_{- 0.044}$ &       1.15 & $  386.08^{+ 336.71}_{-   0.34}$ & $ 3.43^{+ 2.196}_{- 0.001}$ &      10.60 \\
       N 3031 & $ 9840.04^{+ 116.91}_{- 117.17}$ & $ 0.36^{+ 0.006}_{- 0.006}$ &       5.12 & $    2.12^{+   0.12}_{-   0.12}$ & $ 3.34^{+ 0.024}_{- 0.023}$ &      20.30 \\
       N 3198 & $ 2830.13^{+  25.46}_{-  25.54}$ & $ 0.54^{+ 0.003}_{- 0.003}$ &       7.77 & $ 1531.59^{+  15.37}_{-  15.42}$ & $ 0.69^{+ 0.003}_{- 0.003}$ &       9.54 \\
       N 3521 & $  195.76^{+  10.31}_{-  10.44}$ & $ 1.42^{+ 0.017}_{- 0.016}$ &       5.82 & $   48.00^{+   2.85}_{-   2.89}$ & $ 1.83^{+ 0.019}_{- 0.018}$ &       6.84 \\
       N 3621 & $ 3728.60^{+  28.09}_{-  28.17}$ & $ 0.50^{+ 0.003}_{- 0.003}$ &       0.87 & $ 1880.38^{+  16.30}_{-  16.34}$ & $ 0.70^{+ 0.003}_{- 0.003}$ &       1.24 \\
       N 4736 & $ 8377.42^{+ 176.93}_{- 177.79}$ & $ <0.002$ &       5.76 & $ 4216.82^{+ 176.24}_{- 176.98}$ & $  <0.003$ &       4.11 \\
       N 5055 & $  322.28^{+   5.04}_{-   5.06}$ & $ 1.16^{+ 0.004}_{- 0.004}$ &       4.96 & $   11.81^{+   0.24}_{-   0.24}$ & $ 2.01^{+ 0.000}_{- 0.005}$ &      13.50 \\
       N 6946 & $ 2005.70^{+  20.02}_{-  20.07}$ & $ 0.92^{+ 0.004}_{- 0.004}$ &       1.25 & $   57.86^{+   1.00}_{-   1.00}$ & $ 2.11^{+ 0.007}_{- 0.007}$ &       2.47 \\
       N 7331 & $  876.99^{+  11.11}_{-  11.15}$ & $ 1.27^{+ 0.004}_{- 0.004}$ &       8.21 & $   80.47^{+   1.26}_{-   1.27}$ & $ 1.99^{+ 0.005}_{- 0.005}$ &      26.00 \\
       N 7793 & $ 2850.27^{+  35.74}_{-  35.91}$ & $ 0.69^{+ 0.009}_{- 0.009}$ &       4.48 & $ 1884.05^{+  26.61}_{-  26.74}$ & $ 0.92^{+ 0.009}_{- 0.009}$ &       5.31 \\
        N 925 & $  100.24^{+   2.31}_{-   2.32}$ & $ 1.83^{+ 0.010}_{- 0.010}$ &       1.35 & $   20.43^{+   0.60}_{-   0.60}$ & $ 2.41^{+ 0.012}_{- 0.012}$ &       2.30 \\
    \end{tabular}
\caption{\footnotesize{Fitted values for the Kroupa and diet-Salpeter stellar mass models with a power-law ($r^n$) dependence on the Rindler acceleration. In
column (2) and (5) we present the result for the Rindler acceleration (in units $\frac{m^2}{{\rm s}^2 kpc^n}$), in column (3) and (6) we present the power-law value of $r^n$.}}
 \label{tab:things3}
\end{table}

\newpage

\newpage

\newpage

\subsection{Fitting the standard and power-law  generalized Rindler model with the Free $\gs$ stellar model} \label{fitting_n1_n}

\begin{figure}
  \includegraphics[width=13cm]{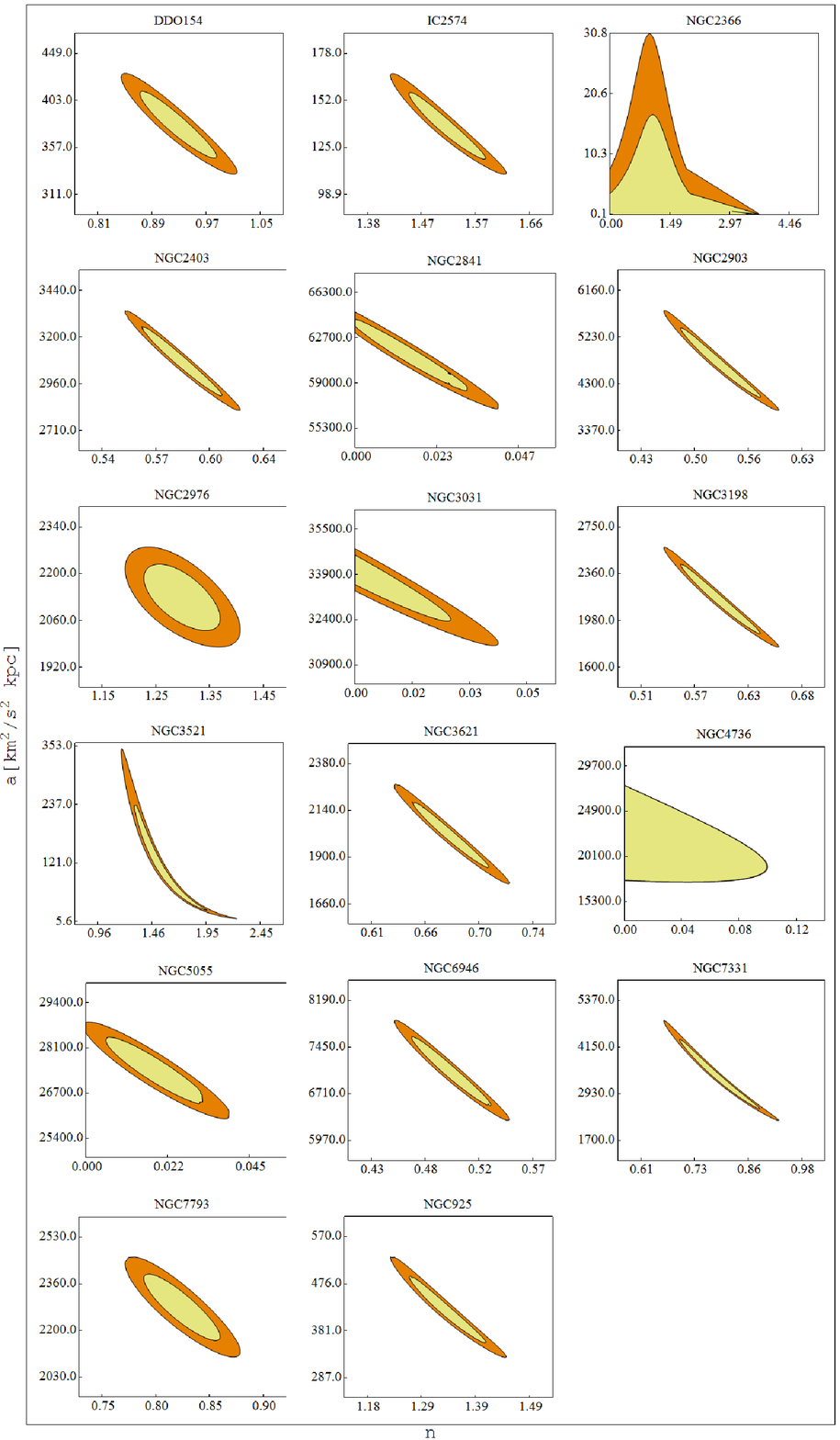}\\
  \caption{\footnotesize{2D likelihood contours between the $a$ and $n$, corresponding to $1\sigma$ and $2\sigma$, for the free mass model.}}
  \label{tab:Free}
\end{figure}

Our results above show that for some galaxies the estimated parameters and/or their associated $\chi^{2}_{\rm red}$ present large discrepancies when we change 
the stellar mass model (Kroupa and diet-Salpeter), see e.g. galaxies N 3031 and N 6946 in Table \ref{tab:things3}.  Indeed, such results depend on the understanding 
on the stellar/baryonic physics in galaxies, and there is an extensive research in the field \cite{Salpeter:1955it,Bottema:1997qe,Kroupa:2000iv,Bell:2000jt}. Because 
we do not  completely comprehend the baryonic physics behind the galaxies we include the stellar ``free mass model"  
in which the parameter $\gs$ is let free and get determined as a best fit, together with the Rindler parameters.  In the few cases when a  bulge was present we treated 
the bulge mass ($M_B$) as a free parameter too, as described in Ref. \cite{deBlok:2008wp}. Again, we started with the standard Rindler exponent ($n=1$) and later
considered $n$ free.  The results are shown in Table \ref{tab:things4}.   The standard Rindler model yields values for $a$ (in cm/s$^2$) varying from less than
$0.1$ for N 4736 to $ 14.80^{+ 0.02}_{- 0.95}$ for N 3521, a difference of
2 orders of magnitude, and in addition, the $\chi^2$ test does not render satisfactory results for some of the galaxies.  Then, we considered the generalized 
Rindler model to obtain the ``acceleration"
parameter ranging from $< 10$ for N 2366 to $ 19022^{+ 153}_{- 154}$ for N 4736, accounting for a 3 orders of magnitude difference.    On the other hand, the 
power-law exponent varies
from   $<0.002$ for N 3031 to $ 1.52^{+ 0.009}_{- 0.009}$ for I 2574, a difference of 2 orders of magnitude.  In Fig. \ref{tab:Free} we present contour plots of the 
Rindler parameters $(a,n)$  for 1$\sigma$ and 2$\sigma$ for the free mass model.

By comparing both fits ($n=1$ vs $n \neq 1$), one observes that both the 
 $\chi^{2}_{\rm red}$ and Rindler acceleration values substantially change. The generalized model achieved a better fit than the standard Rindler model for all the galaxies. 
 Now, comparing the best fits of Tables \ref{tab:things3} and  \ref{tab:things4}, i.e., generalized Rindler model with Kroupa vs free mass, the
results are better in 14 cases (out of 17) for the free mass model.  The price paid is that the free stellar model in the generalized Rindler model has a variation
of the exponent ($n$) of 3 orders of magnitude against a variation of {\it only} 2 orders of magnitude in the Kroupa model.

In a very recent work a similar model is considered \cite{Lin:2012zh} to fit the same eight (out of our seventeen) galaxies fixing $n=1$ and letting $\gs$ to be free. The authors
fitted the following galaxies: N 2403, N 2841, N 2903, N 3198, N 3521, N 5055, N 7331, and D 154. However, they have not taken into account color gradients nor bulges, as we had, so their conclusions are not expected to be identical to ours. They concluded that six of these galaxies tend fit
well to the data and that there is a preferred Rindler acceleration parameter of around $a \approx 3.0 \times 10^{-9} \, {\rm cm}/{\rm s}^2$; they later fixed this acceleration parameter and found acceptable fits for five galaxies, and furthermore, an additional free parameter let them fit two more galaxies.  When we analyze these galaxies in our results in Table \ref{tab:things4} we reach similar conclusions on the fits (except for N 3521) and to a convergence to a similar Rindler acceleration within $1 \sigma$ confidence level.   However, when one adds more galaxies to the analysis the spread in the acceleration blows up, as shown in
Table \ref{tab:things4}.

\setlength{\extrarowheight}{3pt}
\begin{table}[h]
  \centering
    \begin{tabular}{r|rrrrr|rrrrr}
    \multicolumn{ 11}{c}{Free mass model} \\
               &                                       \multicolumn{ 5}{c}{$n=1$} &                                      \multicolumn{ 5}{c}{$n \neq 1$} \\
        Galaxy &$ a$ & $a$ $[\frac{\rm{cm}}{{\rm s}^2}]$ & $M_{D} $& $ M_{B}$  &  $\chi^{2}_{\rm red}$ & $a$ &    $ n$  & $M_{D} $& $ M_{B} $   &    $\chi^{2}_{\rm red}$ \\
        \hline
        D 154 & $  332.91^{+   5.42}_{-   5.50}$ & $  1.08^{+ 0.02}_{- 0.02}$ & $ 7.96^{+ 7.06}_{- 7.07}$ &            &       1.78 & $  378.75^{+   6.01}_{-   6.05}$ & $ 0.93^{+ 0.010}_{- 0.010}$ & $ 7.80^{+ 7.06}_{- 7.06}$ &            &       1.75 \\
        I 2574 & $  365.24^{+   6.89}_{-   6.95}$ & $  1.18^{+ 0.02}_{- 0.02}$ &   $ < 1 $ &            &       3.94 & $  136.85^{+   2.63}_{-   2.65}$ & $ 1.52^{+ 0.009}_{- 0.009}$ & $ 8.26^{+ 7.76}_{- 7.77}$ &            &       1.03 \\
       N 2366 & $  176.84^{+  16.41}_{-  16.69}$ & $  0.57^{+ 0.05}_{- 0.05}$ & $ 9.00^{+ 7.12}_{- 8.69}$ &            &       2.30 & $    < 10 $ & $ 1.11^{+1.00}_{- 1.00}$ & $ 9.31^{+ 9.26}_{- 7.92}$ &            &       1.58 \\
       N 2403 & $  797.22^{+  97.65}_{-   0.32}$ & $  2.58^{+ 0.32}_{- 0.30}$ & $ 10.2^{+ 9.83}_{- 6.36}$ & $ 6.77^{+ 5.80}_{- 5.72}$ &       4.88 & $ 3070.1^{+  16.1}_{-  16.1}$ & $ 0.59^{+ 0.002}_{- 0.002}$ & $ 9.85^{+ 8.07}_{- 8.07}$ & $ 7.70^{+ 6.65}_{- 6.65}$ &       0.75 \\
       N 2841 & $ 1182.6^{+  16.57}_{-  16.73}$ & $  3.83^{+ 0.05}_{- 0.05}$ & $ 11.4^{+ 9.08}_{- 9.08}$ &            &       1.08 & $ 61227^{+ 358}_{- 359}$ & $ 0.02^{+ 0.002}_{- 0.002}$ & $ 11.0^{+ 9.08}_{- 9.08}$ &            &       0.14 \\
       N 2903 & $  965.84^{+  12.81}_{-  12.87}$ & $  3.13^{+ 0.04}_{- 0.04}$ & $ 10.8^{+ 8.74}_{- 8.75}$ & $ 7.43^{+ 6.39}_{- 6.47}$ &       2.47 & $ 4686.8^{+  52.1}_{-  52.3}$ & $ 0.53^{+ 0.004}_{- 0.004}$ & $ 10.7^{+ 8.74}_{- 8.75}$ & $ 7.70^{+ 6.71}_{- 6.77}$ &       1.82 \\
       N 2976 & $ 2646.5^{+  99.93}_{-  21.12}$ & $  8.58^{+ 0.32}_{- 0.07}$ & $5.86^{+6.95}_{-4.95}$ &            &       2.32 & $ 2130.7^{+  41.2}_{-  41.6}$ & $ 1.30^{+ 0.030}_{- 0.030}$ & $ 8.38^{+ 7.59}_{- 7.59}$ &            &       0.99 \\
       N 3031 & $ 1602.5^{+  31.69}_{-  31.80}$ & $  5.19^{+ 0.10}_{- 0.10}$ & $ 10.8^{+ 8.50}_{- 8.50}$ & $ 7.83^{+ 6.79}_{- 6.78}$ &       5.93 & $ 34102^{+ 233}_{- 228}$ &  $<0.002$ & $ 10.5^{+ 8.50}_{- 8.51}$ & $ 5.86^{+ 4.88}_{- 4.89}$ &       4.86 \\
       N 3198 & $  503.00^{+   5.72}_{-   5.74}$ & $  1.63^{+ 0.02}_{- 0.02}$ & $ 10.5^{+ 8.62}_{- 8.62}$ & $ 7.96^{+ 6.95}_{- 6.97}$ &       5.11 & $ 2144.6^{+  21.0}_{-  21.0}$ & $ 0.60^{+ 0.003}_{- 0.003}$ & $ 10.4^{+ 8.62}_{- 8.62}$ & $ 8.65^{+ 7.64}_{- 7.69}$ &       3.96 \\
       N 3521 & $ 4580.0^{+   7.67}_{- 294}$ & $ 14.80^{+ 0.02}_{- 0.95}$ & $9.00^{+6.88}_{-6.45}$ &            &      63.10 & $  120.08^{+   6.75}_{-   8.96}$ & $ 1.51^{+ 0.017}_{- 0.023}$ & $ 11.0^{+ 8.96}_{- 9.05}$ &            &       1.78 \\
       N 3621 & $  689.06^{+   4.84}_{-   9.49}$ & $  2.23^{+ 0.02}_{- 0.03}$ & $ 10.4^{+ 8.39}_{- 8.32}$ &            &       0.67 & $ 2015.0^{+  17.3}_{-  17.4}$ & $ 0.67^{+ 0.003}_{- 0.003}$ & $ 10.3^{+ 8.35}_{- 8.35}$ &            &       0.56 \\
       N 4736 & $    < 10 $ & $    < 0.1 $ & $ 10.5^{+ 8.44}_{- 8.44}$ & $ 7.27^{+ 6.34}_{- 6.24}$ &      53.0 & $ 19022^{+ 153}_{- 154}$ & $ 0.10^{+ 0.001}_{- 0.001}$ & $7.08^{+7.90}_{-5.90}$ & $ 6.69^{+ 5.77}_{- 5.63}$ &      32.1 \\
       N 5055 & $  556.69^{+   8.75}_{-   8.82}$ & $  1.80^{+ 0.03}_{- 0.03}$ & $ 11.0^{+ 8.76}_{- 8.77}$ &      $<1$ &       9.30 & $ 27370^{+ 175}_{- 175}$ & $ 0.02^{+ 0.002}_{- 0.002}$ & $ 10.5^{+ 8.76}_{- 8.76}$ & $ 7.70^{+ 6.62}_{- 6.75}$ &       0.86 \\
       N 6946 & $ 1378.7^{+  16.64}_{-  16.30}$ & $  4.47^{+ 0.05}_{- 0.05}$ & $ 10.7^{+ 8.55}_{- 8.54}$ &      $<1$ &       1.48 & $ 7052.8^{+  53.5}_{-  53.6}$ & $ 0.50^{+ 0.003}_{- 0.003}$ & $ 10.6^{+ 8.54}_{- 8.54}$ & $ 3.70^{+ 2.67}_{- 2.70}$ &       1.22 \\
       N 7331 & $ 1707.9^{+  24.65}_{-  24.73}$ & $  5.53^{+ 0.08}_{- 0.08}$ & $ 11.0^{+ 9.01}_{- 9.01}$ & $ 12.2^{+ 11.2}_{- 11.1}$ &       0.26 & $ 3348.2^{+  45.7}_{-  45.8}$ & $ 0.79^{+ 0.005}_{- 0.005}$ & $ 10.9^{+ 9.01}_{- 9.01}$ & $ 7.70^{+ 6.67}_{- 6.77}$ &       0.24 \\
       N 7793 & $ 1523.1^{+  23.79}_{-  23.79}$ & $  4.94^{+ 0.08}_{- 0.08}$ & $ 9.47^{+ 7.85}_{- 7.85}$ &            &       4.88 & $ 2277.8^{+  30.2}_{-  30.4}$ & $ 0.83^{+ 0.009}_{- 0.009}$ & $ 9.31^{+ 7.85}_{- 7.85}$ &            &       4.61 \\
        N 925 & $  803.09^{+   0.42}_{- 450}$ & $  2.60^{+ 0.01}_{- 1.46}$ & $ 7.55^{+ 6.88}_{-6.88}$ &            &       4.07 & $  420.11^{+   7.12}_{-   7.17}$ & $ 1.33^{+ 0.008}_{- 0.008}$ & $ 9.47^{+ 8.34}_{- 8.34}$ &            &       1.12 \\
    \end{tabular}
  \caption{\footnotesize{Fits for free mass model. In columns (2), (3) and (7) we present the Rindler acceleration in units of
  $\frac{km^2}{{\rm s}^2 kpc}$, $\rm{cm}/{\rm s}^2 \times 10^{-9}$ and $\frac{m^2}{{\rm s}^2 kpc^n}$, respectively. The stellar disk ($M_D$)
  and bulge ($M_B$) masses are given in log$_{10}$ solar mass units. }}
  \label{tab:things4}
\end{table}

\newpage

\subsection{Comparison of Rindler models with dark matter profiles} \label{dm_comparison}

In order to compare the above results with those of standard dark matter profiles we include the fittings of the NFW \cite{Navarro:1995iw,Navarro:1996gj}
and Burkert  \cite{Burkert:1995yz} profiles for the Kroupa and diet-Salpeter stellar mass models.
The results are shown in Tables   \ref{tab:things6} and  \ref{tab:things7}, which were extracted from our previous  work  \cite{Mastache:2011cn}, where we employed
the same galaxies and the fits were computed with the same technique as in the present work.  In general, as found
by many authors cored profiles fit better than cuspy profiles to the type of galaxies study here, see e.g.
Refs \cite{SpGiHa05,KuMcBl08,Salucci:2010pz,Gentile:2004tb,Salucci:2007tm,deBlok:2008wp,delaMacorra:2011df,Mastache:2011cn}.   This is evident from the
lower $\chi^{2}_{ \rm red}$  values for Burkert's profile in comparison to NFW's for most of the galaxies for both stellar mass models.

To compare among the different models we constructed Table \ref{tab:things5} with the $\chi^{2}_{ \rm red}$ values for  NFW, Burkert,  standard
Rindler with $n=1$, and generalized Rindler $n \neq 1$, for both Kroupa and
diet-Salpeter stellar mass models.   The results are as follows:

\begin{itemize}

\item As expected, the Rindler model with two free parameters ($a$, $n$) fits better than the model with a single parameter ($a$, $n=1$)  for both Kroupa and diet-Salpeter stellar mass models, for all galaxies.

\item The standard Rindler model ($n=1$) does fit worst than NFW's and Burkert's profiles for the Kroupa mass model, and the same trend is observed for the diet-Salpeter stellar model. In this later case,  the standard Rindler model fits better than NFW and Burkert only for one galaxy (N 3521), and it fits better than Burkert for one galaxy  (N 3621), and it fits 
equally as well as NFW for two galaxies (N 6946 and N 7331).

\item Considering  the generalized Rindler model ($n \neq 1$) for the Kroupa stellar mass model the fits are better than the standard Rindler model, since one now has 2 degrees of freedom.   This model fits better than both NFW and Burkert models for four galaxies (N 2841, N 2976, N 3521, and N 3621) and, in addition, it does fit better than NFW for four galaxies (IC2574, N 5055, N 7331, and N 925) and better than Burkert for one galaxy (N 7793) and it fits equally well as one galaxy (N 3031).  In all other cases, however, 20 of the 34 possibilities, NFW and Burkert profiles for the Kroupa stellar mass model fit  better than the generalized Rindler model: The NFW profile fits better for 9 galaxies (out of 17) than the generalized  Rindler model and Burkert's profile achieves a better fit for 11 galaxies (out of 17).

\item The generalized Rindler model ($n \neq 1$) for the diet-Salpeter stellar mass model results are slightly better than the Kroupa's model:  The model fits better than both NFW and Burkert models for five galaxies (N 3031, N 3521, N 3621, N 6946, and N 925) and, in 
addition, it fits better than NFW for three galaxies (IC2574, N 5055, and N 7331), but it does not fit better for any other galaxy with Burkert's 
profile.   Taking into account all the other cases, however, 21 of the 34 possibilities, NFW and Burkert profiles for the diet-Salpeter stellar 
mass model  fit  better than the generalized Rindler model: The NFW profile fits better for 9 galaxies (out of 17)   and Burkert profile 
achieves a better fit for 12 galaxies (out of 17) than the generalized  Rindler model.

\end{itemize}

\begin{figure}
  \centering
    \subfloat[\footnotesize{N 2841}]{ \includegraphics[width=0.30\textwidth]{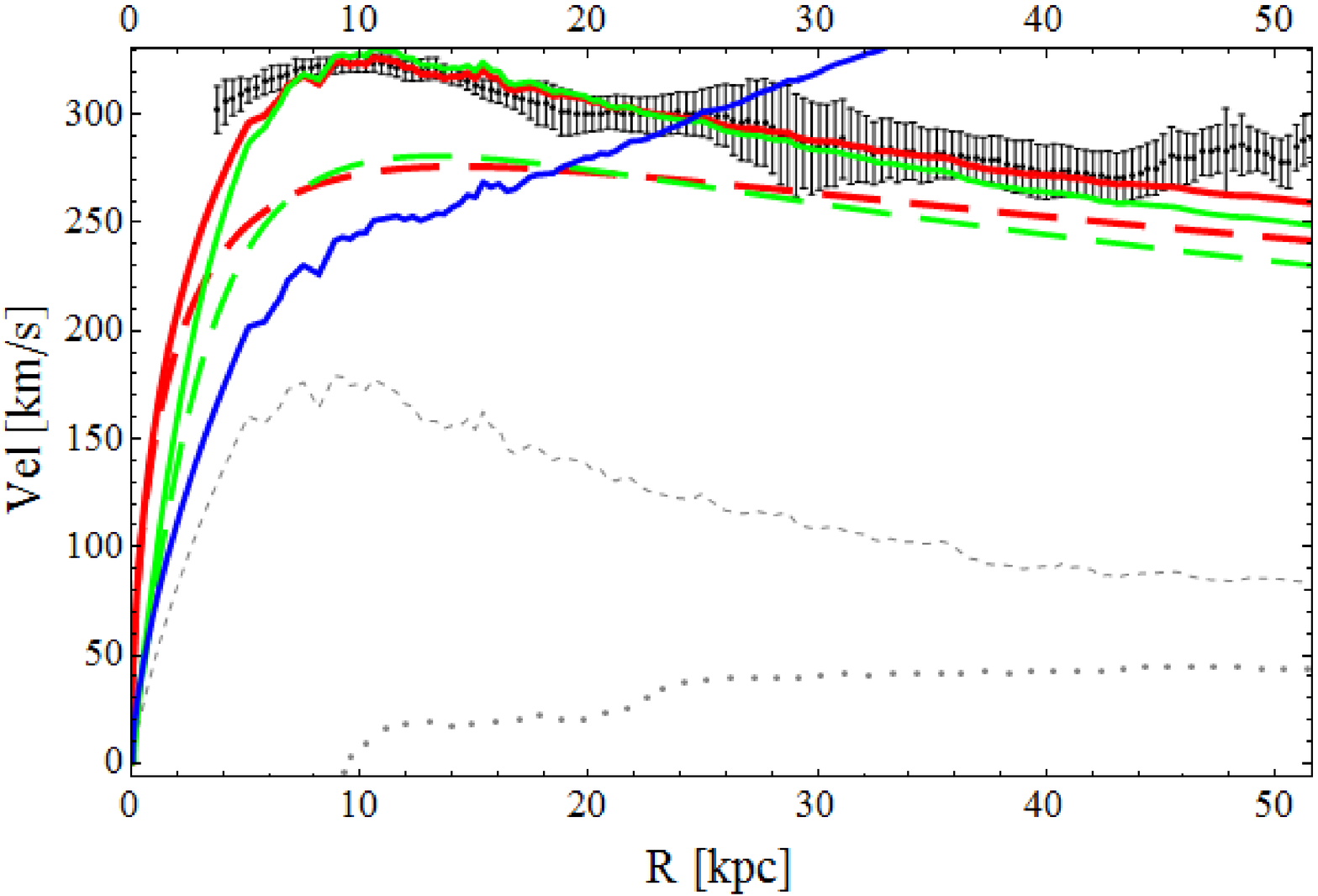}}
    \subfloat[\footnotesize{N 3621}]{ \includegraphics[width=0.30\textwidth]{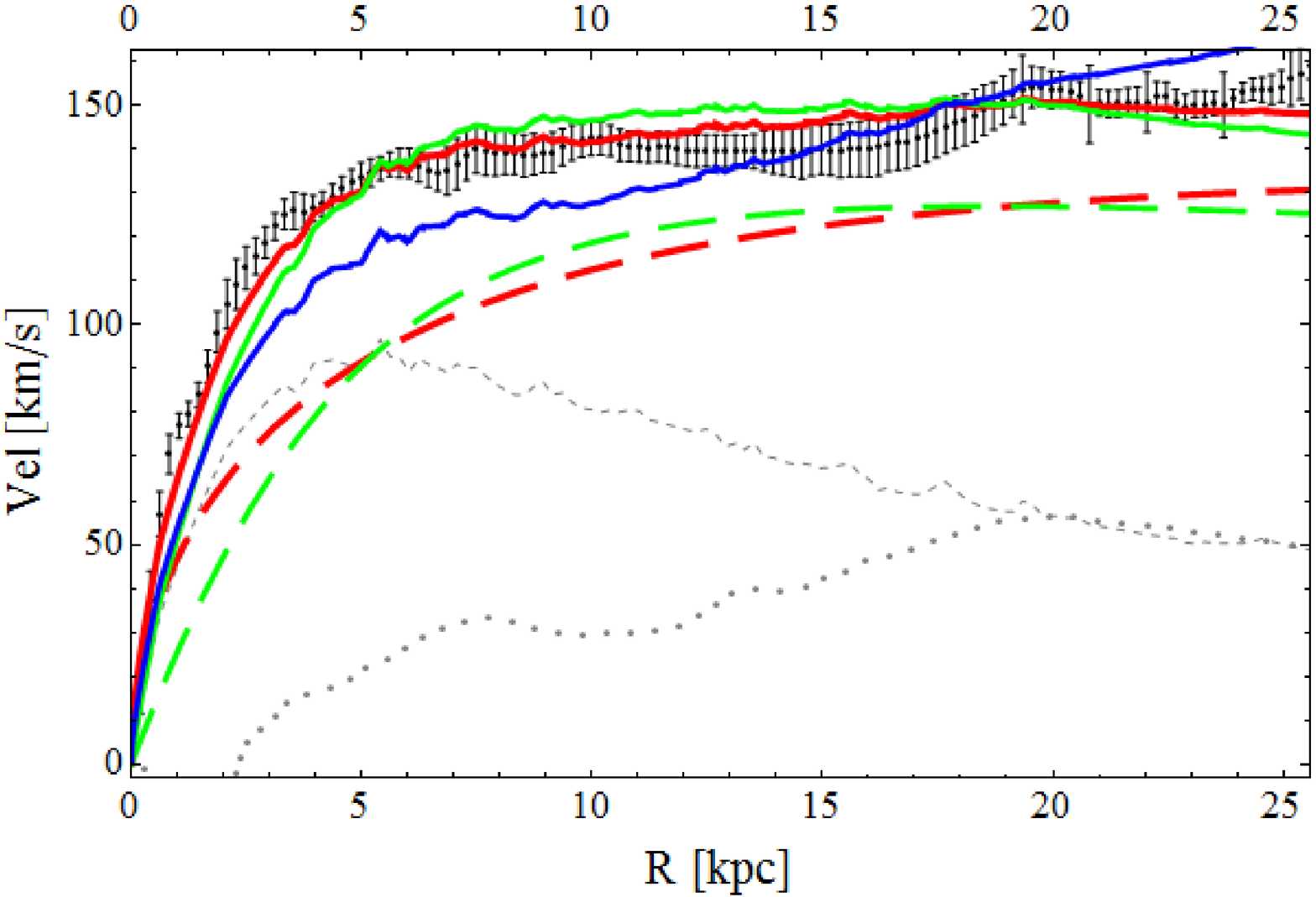}}
    \subfloat[\footnotesize{N 6946}]{ \includegraphics[width=0.30\textwidth]{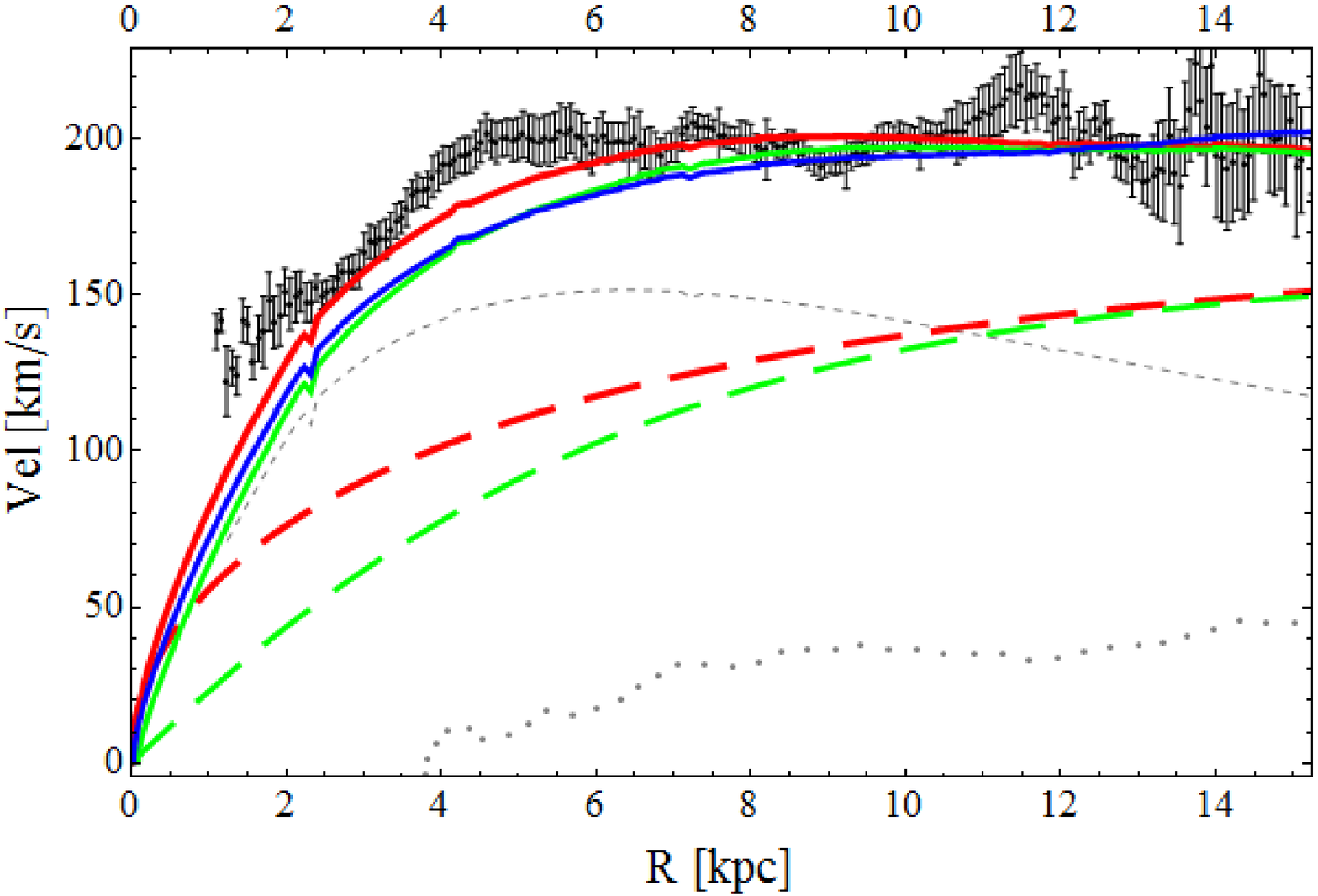}}
    \caption{\footnotesize{
    Rotation curve for the galaxies (a) N 2841, (b) N 3621, and (c) N 6946. We explicitly plot the contribution of the gas (dotted, black), the Kroupa star model (dashed, black), the red/green plots are the NFW/Burkert DM halo profile and its total contribution (dashed and thick lines, respectively), and finally the blue line is the total rotation curve of  the standard ($n=1$) Rindler acceleration model.
    }}
    \label{fig:rctotal}
\end{figure}

For completeness,  in Fig. (\ref{fig:rctotal}) we show rotation curves of three galaxies (N 2841, N 3621, and N 6946) as examples of typical fits 
for the standard ($n=1$) Rindler model and its comparison with NFW and Burkert profiles.  For some of the 
galaxies (e.g. N 2841), the rotation curve linear term,  given by Eq. \ref{aRind}, overestimates the external rotational 
curve given rise to bad fits, a point that was warned in  Ref. \cite{Grumiller:2010bz} and also discussed in Ref. \cite{Lin:2012zh}.  On the contrary, 
it is argued, e.g. in Ref. \cite{Salucci:2007tm}, that rotation curves tend to slowly decrease after a few optical radius.  

\setlength{\extrarowheight}{3pt}
\begin{table}[h]
  \centering
    \begin{tabular}{r|rrr|rrr}
                                                                 \multicolumn{ 7}{c}{Kroupa} \\
               \hline
               &             \multicolumn{ 3}{c|}{NFW} &         \multicolumn{ 3}{c}{Burkert} \\
        Galaxy &     $r_s$ & $\log \rho_0$  & $\chi^{2}_{\rm red}$ &     $r_s$ & $\log \rho_0$  & $\chi^{2}_{ \rm red}$ \\
        \hline
        D 154 & $15.14^{+0.17}_{-0.18}$ & $6.1^{+4.28}_{-4.28}$ &       1.06 & $2.47^{+0.02}_{-0.02}$ & $7.43^{+5.61}_{-5.6}$ &       0.43 \\
        I  2574 &   $>10^6$ & $ <0.1$ &        2.4 & $14.74^{+0.57}_{-0.54}$ & $6.62^{+4.96}_{-4.95}$ &       0.69 \\
       N 2366 & $15.47^{+0.58}_{-0.57}$ & $6.07^{+4.75}_{-4.74}$ &          3 & $1.75^{+0.04}_{-0.05}$ & $7.72^{+6.37}_{-6.37}$ &       1.34 \\
       N 2403 &   $10.38^{+0.03}_{-0.03}$ & $7.14^{+4.8}_{-4.8}$ &      0.82 &   $3.86^{+0.01}_{-0.01}$ & $7.98^{+5.63}_{-5.63}$ &  1.6 \\
       N 2841 & $6.67^{+0.02}_{-0.02}$ & $8.18^{+5.89}_{-5.89}$ &       1.29 & $4.08^{+0.01}_{-0.01}$ & $8.62^{+6.34}_{-6.33}$ &       3.02 \\
       N 2903 & $4.92^{+0.02}_{-0.02}$ & $8.12^{+5.94}_{-5.94}$ &       3.43 & $3.06^{+0.01}_{-0.01}$ & $8.55^{+6.37}_{-6.45}$ & 2.01 \\
       N 2976 &   $>10^5$ & $1.77^{+0.46}_{-0.15}$ &        6.3 &   $>10^4$ & $7.69^{+6.19}_{-6.2}$ &          1 \\
       N 3031 & $8.35^{+0.06}_{-0.06}$ & $7.48^{+5.56}_{-5.55}$ &       4.96 & $3.5^{+0.02}_{-0.03}$ & $8.22^{+6.29}_{-6.3}$ &       5.12 \\
       N 3198 & $20.95^{+0.11}_{-0.11}$ & $6.54^{+4.5}_{-4.47}$ &       4.75 & $8.1^{+0.04}_{-0.04}$ & $7.34^{+5.29}_{-5.28}$ &       2.49 \\
       N 3521 &   $>10^6$ & $1.03^{+-0.27}_{--0.28}$ &       5.66 & $15.70^{+0.58}_{-0.56}$ & $6.93^{+5.61}_{-5.61}$ &       5.21 \\
       N 3621 & $17.1^{+0.08}_{-0.08}$ & $6.72^{+4.6}_{-4.6}$ &       1.45 & $5.77^{+0.02}_{-0.02}$ & $7.63^{+5.51}_{-5.51}$ &       5.59 \\
       N 4736 & $0.23^{+0.02}_{-0.02}$ & $10.44^{+8.74}_{-8.74}$ &       1.34 & $0.22^{+0.05}_{-0.05}$ & $10.45^{+8.76}_{-8.76}$ &        1.3 \\
       N 5055 &   $>10^6$ & $ <2 $ &       5.09 & $23.54^{+0.27}_{-0.26}$ & $6.6^{+4.79}_{-4.8}$ &       4.11 \\
       N 6946 & $96.74^{+0.84}_{-0.85}$ & $5.87^{+3.87}_{-3.87}$ &       1.21 & $8.12^{+0.06}_{-0.06}$ & $7.52^{+5.52}_{-5.52}$ &       1.14 \\
       N 7331 &   $>10^6$ &      $<2$ &       8.63 &   $17.4^{+0.19}_{-0.19}$ & $7.23^{+5.33}_{-5.34}$ & 7.06 \\
       N 7793 & $8.62^{+0.08}_{-0.08}$ & $7.15^{+5.26}_{-5.25}$ &       4.07 & $1.88^{+0.01}_{-0.02}$ & $8.35^{+6.44}_{-6.45}$ &       7.88 \\
        N 925 &   $>10^5$ & $1.95^{+0.34}_{-0.32}$ &       3.68 & $21.15^{+1.02}_{-0.95}$ & $6.77^{+5.14}_{-5.13}$ &       1.19 \\
    \end{tabular}
  \caption{\footnotesize{Fitted parameter values for the NFW and Burkert profiles with a Kroupa mass stellar model.
  }}
  \label{tab:things6}
\end{table}

\setlength{\extrarowheight}{3pt}
\begin{table}[h]
  \centering
    \begin{tabular}{r|rrr|rrr}
                                                          \multicolumn{ 7}{c}{diet-Salpeter} \\
                                                          \hline
               &             \multicolumn{ 3}{c|}{NFW} &          \multicolumn{ 3}{c}{Burkert} \\
        Galaxy &     $r_s$ & $\log \rho_0$  & $\chi^{2}_{\rm red}$ &     $r_s$ & $\log \rho_0$  & $\chi^{2}_{\rm red}$ \\
        \hline
        D 154 & $15.95^{+0.19}_{-0.19}$ & $6.07^{+4.25}_{-4.25}$ &       1.09 & $2.53^{+0.03}_{-0.03}$ & $7.41^{+5.59}_{-5.59}$ &       0.39 \\
        I 2574 &   $>10^5$ &      $<2$ &       5.49 & $20.82^{+1.27}_{-1.14}$ & $6.51^{+4.89}_{-4.89}$ &       1.16 \\
       N 2366 & $15.99^{+0.63}_{-0.62}$ & $6.03^{+4.74}_{-4.71}$ &       2.94 & $1.78^{+0.05}_{-0.05}$ & $7.69^{+6.36}_{-6.35}$ &       1.34 \\
       N 2403 & $12.53^{+0.03}_{-0.03}$ & $6.99^{+4.65}_{-4.65}$ & 1.08 & $5.71^{+0.02}_{-0.02}$ & $7.63^{+5.33}_{-5.32}$ &       1.37 \\
       N 2841 & $5.84^{+0.43}_{-0.43}$ & $8.37^{+6.51}_{-6.51}$ &  0.49 & $3.85^{+0.03}_{-0.03}$ & $8.75^{+7.02}_{-7.01}$ &    0.93 \\
       N 2903 & $4.81^{+0.01}_{-0.01}$ & $8.15^{+6.04}_{-5.96}$ &  3.73 & $2.85^{+0.01}_{-0.01}$ & $8.63^{+6.44}_{-6.43}$ & 1.23 \\
       N 2976 &   $>10^5$ & $2.41^{+1.22}_{-1.22}$ &      10.49 &   $>10^4$ & $7.46^{+6.41}_{-5.95}$ &       5.11 \\
       N 3031 &   $>10^6$ &      $<2$ &      22.76 &   $1.90^{+0.01}_{-0.01}$ & $9.06^{+7.82}_{-7.81}$ &    4.37 \\
       N 3198 & $41.1^{+0.27}_{-0.27}$ & $6.01^{+4.01}_{-4.01}$ &  7.86 & $9.74^{+0.05}_{-0.05}$ & $7.17^{+5.16}_{-5.13}$ & 0.93 \\
       N 3521 &   $>10^6$ &      $<2$ & 12.5 &   $162^{+135}_{-50.1}$ & $6.22^{+5.02}_{-5.01}$ & 11.5 \\
       N 3621 & $43.81^{+0.27}_{-0.27}$ & $6.03^{+3.98}_{-3.98}$ &       1.44 & $9.41^{+0.05}_{-0.05}$ & $7.21^{+5.15}_{-5.17}$ &       3.64 \\
       N 4736 & $0.07^{+0.01}_{-0.01}$ & $11.54^{+10.11}_{-10.16}$ &        1.3 & $0.08^{+0.01}_{-0.01}$ & $11.34^{+9.91}_{-9.96}$ &       1.29 \\
       N 5055 &   $>10^6$ & $ <1 $ &      15.74 & $2.52^{+0.03}_{-0.03}$ & $8.79^{+6.89}_{-6.89}$ & 3.47 \\
       N 6946 &   $>10^5$ & $2.89^{+1.15}_{-1.14}$ &       4.42 & $309.22^{+322.11}_{-104.69}$ & $6.65^{+4.89}_{-4.9}$ &       2.49 \\
       N 7331 &   $>10^6$ &      $<2$ & 29.6 &   $27.58^{+0.49}_{-0.48}$ & $7.00^{+5.15}_{-5.14}$ & 15.57 \\
       N 7793 & $30.08^{+0.36}_{-0.36}$ & $6.4^{+4.55}_{-4.55}$ &       5.01 & $3.09^{+0.03}_{-0.03}$ & $7.98^{+6.14}_{-6.13}$ &        4.3 \\
        N 925 &   $>10^6$ & $1.23^{+0.26}_{-0.27}$ &       5.64 & $61.11^{+18.46}_{-8.61}$ & $6.54^{+4.78}_{-5.23}$ &       2.54 \\
    \end{tabular}
  \caption{\footnotesize{Fitted parameter values for the NFW and Burkert profiles with a diet-Salpeter mass stellar model.
  }}
  \label{tab:things7}
\end{table}

\setlength{\extrarowheight}{3pt}
\begin{table}[h]
  \centering
    \begin{tabular}{r|rrrr|rrrr}
                                                                                 \multicolumn{ 9}{c}{\bf Goodness-of-fit comparison table} \\
                                                                                 \hline
               &                       \multicolumn{ 4}{c|}{Kroupa} &                \multicolumn{ 4}{c}{diet-Salpeter} \\
        Galaxy &        NFW &    Burkert & Rindler  & Rindler  &        NFW &    Burkert & Rindler  & Rindler \\
               &            &            &   ($n=1$)    &  ($n\neq1$)&            &            &  ($n=1$)     &  $(n\neq1)$ \\
        \hline
        D 154 &       1.06 &       0.43 &       2.10 &       1.77 &       1.09 &       0.39 &       2.03 &       1.75 \\
        I 2574 &       2.40 &       0.69 &       4.74 &       0.98 &       5.49 &       1.16 &       5.53 &       1.32 \\
       N 2366 &       3.00 &       1.34 &       3.48 &       3.27 &       2.94 &       1.34 &       3.29 &       3.23 \\
       N 2403 &       0.82 &       1.60 &      11.80 &       2.27 &       1.08 &       1.37 &       9.29 &       2.55 \\
       N 2841 &       1.29 &       3.02 &      76.10 &       0.90 &       0.49 &       0.93 &      44.30 &       1.23 \\
       N 2903 &       3.43 &       2.01 &      71.80 &       7.99 &       3.73 &       1.23 &      74.10 &       8.34 \\
       N 2976 &       6.31 &       1.00 &       8.05 &       1.15 &      10.49 &       5.11 &      10.70 &      10.60 \\
       N 3031 &       4.96 &       5.12 &       9.54 &       5.12 &      22.76 &       4.37 &      22.80 &      20.30 \\
       N 3198 &       4.75 &       2.49 &      14.00 &       7.77 &       7.86 &       0.93 &      11.50 &       9.54 \\
       N 3521 &       5.66 &       5.21 &       6.14 &       5.82 &      12.48 &       11.5 &       7.54 &       6.84 \\
       N 3621 &       1.45 &       5.59 &      11.20 &       0.87 &       1.44 &       3.64 &       3.29 &       1.24 \\
       N 4736 &       1.34 &       1.30 &      20.50 &       5.76 &       1.30 &       1.29 &       8.74 &       4.11 \\
       N 5055 &       5.09 &       4.11 &       5.09 &       4.96 &      15.74 &       3.47 &      15.80 &      13.50 \\
       N 6946 &       1.21 &       1.14 &       1.29 &       1.25 &       4.42 &       2.49 &       4.42 &       2.47 \\
       N 7331 &       8.63 &       7.06 &       8.65 &       8.21 &      29.60 &       15.6 &      29.60 &      26.00 \\
       N 7793 &       4.07 &       7.88 &      10.50 &       4.48 &       5.01 &       4.30 &       5.52 &       5.31 \\
        N 925 &       3.68 &       1.19 &       3.74 &       1.35 &       5.64 &       2.54 &       5.72 &       2.30 \\
    \end{tabular}
  \caption{\footnotesize{Summary of the $\chi^2_{\rm red}$ values for the different profiles with the Kroupa and
  diet-Salpeter stellar mass models.   }}
  \label{tab:things5}
\end{table}

As we showed in Refs. \cite{delaMacorra:2011df,Mastache:2011cn}, the BDM profile fits even better than or at least equally as well as Burkert's 
and NFW's for the same galaxies considered here, but  in order to avoid multiple comparisons that may be cumbersome, we have not 
included the BDM profile results here that are  extensively discussed elsewhere \cite{Mastache:2011cn}.  By 
considering alternatives to dark matter profiles one would desire to accomplish a better phenomenology, as the BDM model does, but 
the Rindler modified gravity does not.

\section{Conclusions and final remarks} \label{conclusions}

We have tested the idea that a modification of the Newtonian potential stemming from a Rindler acceleration that modifies gravity, as
proposed by Ref.  \cite{Grumiller:2010bz}, would account for the dark matter content in spiral galaxies. The theory
yields an effective new term in the theoretical rotation curve of the form $v_{T}^2 (r) = \gs v^2_{\star} + v^2_{G} + a |\vec{r}|^n $, where the last term would
replace the contribution of the dark matter profile.

We have made use of the HI data provided by THINGS \cite{Walter:2008wy}, which possess  high resolution velocity fields of rotation curves that are ideal to
test new dark matter profiles  \cite{Mastache:2011cn,delaMacorra:2011df} or, as in the present work, new gravity models.  We have considered the gas component
that is computed by integrating its surface brightness as in the standard Newtonian theory, three stellar mass models (Kroupa, diet-Salpeter, and free mass), and
the standard Rindler model ($n=1$) and a generalized power-law ($n$ free).    We have considered three stellar mass models since the galactic dynamics is 
encoded in the baryonic physics, and this is somewhat unknown. Therefore, the parameter determination of the modified gravity depends on the  
the understanding of the stellar/baryonic physics in galaxies, and to reach conclusions one has to be cautious, through finding best fits and looking for 
convergence of parameters' values. 

The results of the fits are shown in Tables   \ref{tab:things2}, \ref{tab:things3}, and \ref{tab:things4} for three stellar mass models for both Rindler models,
$n=1$ and $n$ free.  We showed, in Sec. \ref{fitting_n=1},
that the fits for $n=1$ are poor, since most of the  $\chi^{2}_{\rm red}$ are bigger than unity.  However, the most important problem is the Rindler acceleration
parameter does not converge to a single value. The computed parameter is in the interval
$ 0.93^{+ 0.01}_{- 0.44} < a <  9.57^{+ 0.06}_{- 0.06}$  in Kroupa's model, to account for a difference of an order of magnitude, whereas
$ 0.62^{+ 0.10}_{- 0.10} < a <  7.79^{+ 0.06}_{- 0.06}$ in diet-Salpeter's model, in a similar fashion as in the previous model. In each stellar mass model
the uncertainties in the Rindler acceleration are small and they could not account for such a big spread in the intervals of $a$.

When the power-law parameter $n$ is set free, see Sec. \ref{fitting_n} and Table \ref{tab:things3}, the fits become better, and by comparing them Kroupa's  did much better in 14 (out of 17) cases
than  diet-Salpeter's.  However, the goodness-of-fit test does not render acceptable results since some galaxies present  very high  $\chi^{2}_{\rm red}$ values. Moreover,  they again suffer from a
broad spread in the determination of the Rindler parameter values, as shown in the contour plots of Fig.  \ref{tab:Kroupa}.  For Kroupa's model the parameters
are $ 77.96^{+   1.74}_{-   1.75} < a < 32294.30^{+ 211.39}_{- 212.01}$, accounting for 2 orders of magnitude in difference and
$0.002 \sim  n < 2.14^{+ 0.045}_{- 0.044}$  that yields a 3 orders of magnitude difference.  For diet-Salpeter parameters the spreads are similar.

When the mass-to-light ratio is set free, see Sec. \ref{fitting_n1_n} and Table \ref{tab:things4}, the fits are better than previous models
when $n$ is a free parameter, but again the spread in parameters is high:
$10 \sim a <19022^{+ 153}_{- 154}$, accounting for a 3 orders of magnitude difference and  $0.002 \sim n < 1.52^{+ 0.009}_{- 0.009}$, a difference of 2 orders of magnitude, see also Fig. \ref{tab:Free}.   A very recent work  \cite{Lin:2012zh} considers  the same problem and using the same eight (of our seventeen) galaxies for
the standard Rindler model ($n=1$), concluding that for six galaxies their results tend to converge to a single Rindler acceleration parameter. In our case, we
observe this evidence too but when one takes into account more galaxies  or other stellar galactic models their conclusions do not hold.

Finally, we have compared our previous results  with two standard dark matter profiles, one cuspy (NFW) and one cored (Burkert), see
Tables \ref{tab:things6}, \ref{tab:things7}, and \ref{tab:things5}.  The goodness-of-fit test favors first Burkert and then NFW profiles over
Rindler's modify gravity.  The tendency is clearer for the standard Rindler ($n=1$) that fits worse than both NFW's and Burkert's profiles for the
Kroupa and diet-Salpeter stellar mass models.  The generalized Rindler model ($n \neq 1$) for the diet-Salpeter stellar mass model results are
slightly better than the Kroupa's model, but still the NFW profile fits better for 9 galaxies (out of 17)  and Burkert profile achieves a better fit for
12 galaxies (out of 17) than the generalized Rindler model.

The overall conclusion is that although the Rindler modified gravity fits are achievable for the considered galaxies, in many cases they show
high  $\chi^{2}_{\rm red}$ values, and a high spread in the Rindler parameters ($a$, $n$) that points for an inconsistent model.  Furthermore, the
standard dark matter profiles (NFW and Burkert) or the alternative BDM model do a better job to fittings of the rotation curves.

\begin{acknowledgments}
We thank Professor Erwin de Blok for providing the observational data of THINGS.  A.M. and J.M.  acknowledge financial
support from Conacyt Project No. 80519 and UNAM PAPIIT Project No. IN100612.
\end{acknowledgments}

\end{document}